\documentclass[aps,pra,10pt,reprint,showpacs,longbibliography]{revtex4-1}

\usepackage{amssymb}
\usepackage{amsmath}
\usepackage{graphicx}% need for figures
\usepackage{hyperref}% use for hypertext links, including those to external documents and URLs
\hypersetup{pdfauthor={Sebastien Berujon and Eric Ziegler},pdftitle={X-Ray Multimodal Tomography Using Speckle-Vector Tracking}}
\DeclareMathOperator*{\argmax}{arg\,max}

\begin{document}
\title{X-Ray Multimodal Tomography Using Speckle-Vector Tracking}
\author{Sebastien Berujon}
\email[]{berujon@esrf.eu}
\affiliation{European Synchrotron Radiation Facility, CS40220, 38043 Grenoble Cedex 9, France}
\author{Eric Ziegler}
\affiliation{European Synchrotron Radiation Facility, CS40220, 38043 Grenoble Cedex 9, France}
\date{\today}

\begin{abstract}
We demonstrate computerized tomography (CT) reconstructions from absorption, phase and dark-field signals obtained from scans acquired when the x-ray probe light is modulated with speckle. Two different interlaced schemes are proposed to reduce the number of sample exposures. First, the already demonstrated x-ray speckle-vector tracking (XSVT) concept for projection imaging allows the three signal CT reconstructions from multiple images per projection. Second, a modified XSVT approach is shown to provide absorption and phase reconstructions, this time from a single image per angular projection. Reconstructions from data obtained at a synchrotron facility emphasize the potential of the approaches for the imaging of complex samples.
\end{abstract}
\maketitle
%\linenumbers

\section{Introduction}
X-ray computed tomography \cite{hounsfield1973} has become an invaluable tool for nondestructive testing and 3D rendering of samples in fields as broad as material sciences, cultural heritage, paleontology and even medical science. In the latter domain, the dose necessary for imaging \emph{in vivo} samples often becomes a critical issue and so must be carefully controlled and maintained below an acceptable hazard threshold \cite{cantril1945,parker1950}. Henceforth, phase and dark-field contrast-imaging methods are regarded as sensible approaches to eventually enhance the rendering quality and quantity of information on soft tissues with a dose level equivalent to or lower than that encountered by using conventional absorption contrast radiography.

Nowadays, a few methods and instruments are available to recover the phase shift induced by a sample exposed to an x-ray light wave. Most techniques use the partial transverse coherence property of an x-ray beam to generate interference between optical waves and enable the calculation of the phase shift. This applies, for instance, to propagation-based methods relying on either the contrast transfer function (holotomography) \cite{cloetens1999} or on the transport-of-intensity equation \cite{paganin2002}. In parallel, the demonstration of grating-based phase-contrast imaging using a laboratory x-ray source promotes the attractiveness of this method \cite{pfeiffer2007}. Therein, a clever setup enables an extended source to be split into a series of smaller-sized sources to permit the generation of constructive interference compatible with shearing interferometry \cite{weitkamp2005,momose2006}. Another method sensitive to the differential phase of an x-ray beam is analyzer-based imaging that employs crystals with a narrow Darwin bandwidth to render different contrast for rays having different propagation orientations \cite{zhao2012}.

Near-field speckle-based methods \cite{cerbino2008,berujon2012prl,morgan2012,berujon2012pra,berujon2015} form a more recent class of x-ray imaging techniques and are proven to be applicable both at synchrotrons and at laboratory sources \cite{zanette2014,zhou2015}. Their attractiveness lies in their low requirements on coherence and in the simplicity of the wave-front modulator, reduced to a simple diffuser such as a piece of sandpaper or a biological filter.

Three main speckle methods relying on pattern correlation in the near field are available today. We purposely leave aside the near-field ptychography technique \cite{stockmar2013} that is based on a different principle (iterative phase recovery) but which often employs a scattering object to improve its efficiency. The first introduced method is x-ray speckle tracking (XST) \cite{berujon2012prl,morgan2012, berujon2015josr}; therein, small subsets of speckle are tracked between two images, one being taken with the sample present in the beam, and the other being used as a reference. The second method requires a slightly more complex arrangement, as the scattering membrane needs to be scanned transversally to the beam with micrometer- or submicrometer-scale steps. This x-ray speckle-scanning (XSS) method \cite{berujon2012pra} provides very high sensitivity at the cost of a number of sample exposures that can quickly rise up to a few dozen. In the last and most recent method \cite{berujon2015}, the scattering object used to generate the static speckle is moved to various transversal positions with respect to the beam while vectors built from speckle images are tracked across reference vectors. We decide to label this method the x-ray speckle vector tracking (XSVT) technique on account of the aspects it shares with the XST technique and the vectors involved in it.

In the first section of this paper, we briefly review and compare these three speckle-processing approaches. Efforts are made to highlight their differences and to demonstrate theoretically, and later experimentally, that XSVT is particularly well suited to performing 3D multimodal computed tomography. Indeed, without any \emph{a priori} assumption on the samples, only a few images per projection are required for multimodal tomography when employing the appropriate scheme. In contrast to previous work \cite{wang2015,zanette2015}, the XSVT approach provides 3D rendering of the dark-field signal, in addition to the absorption and phase signals, whilst the spatial resolution of the imagery is limited only by the detector used.

In the last part, we introduce a mixed approach based on the combination of the XST and XSVT methods. This original scheme permits the extraction of the phase signal with a resolution approaching that of the XSVT technique with a lesser number of exposures than for XSVT alone. Experimentally, the phase-contrast tomography reconstruction obtained from this mixed approach shows that the number of sample exposures necessary is comparable to the case of absorption tomography.

\section{Speckle-processing schemes \label{sec:schemes}}
\begin{figure*}[t]
\includegraphics[width=1\textwidth]{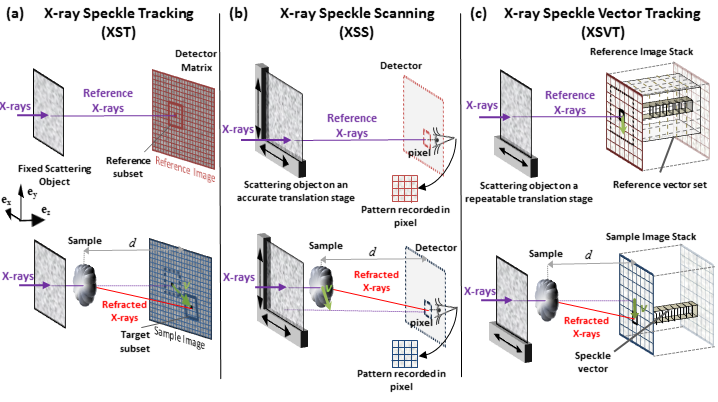}
\caption{Conceptual sketches showing the data recorded and processed within each speckle-processing scheme. (a) X-ray Speckle Tracking. (b) X-ray Speckle Scanning. (c) X-ray Speckle-Vector Tracking. The reference data are drawn in red (top line) and the 'sample' data in blue (bottom line).\label{fig:methods}}
\end{figure*}

We give here a short comparative review of the speckle-processing approaches.

The three speckle schemes available today differ in the kind of data collected, the numerical treatment applied and, for two of them, their intrinsic sensed signals. Nevertheless, the concept shared by all three schemes is the use of cross-correlation operations to track the displacement of a distorted pattern with respect to a reference pattern \cite{pan2009} caused by x-ray refraction through the sample. In all schemes, the speckle pattern used as structured illumination is generated by interference upon the propagation of coherent photons through a scattering object (scatterer) with small random features, often made of sandpaper or of a biological filtering membrane with a known statistical feature size.

To be more precise, the insertion of a sample into an x-ray beam generates a dephasing of the beam waves by the amount $\phi = -k\int \delta(x,y,z) dz$ where $k$ is the wave number and $\delta$ is the optical refractive index of the sample material. Upon propagation along an $\mathbf{e_z}$ axis, this dephasing is responsible for the distortion of an input reference modulation pattern, i.e., here, the speckle pattern. The local speckle shift is equal to $\mathbf{v} = \frac{d}{k} \nabla \phi$ in the basis $(\mathbf{e_x},\mathbf{e_y})$ transverse to the photon-propagation direction. Therein, $\nabla$ is the del operator and $d$ the propagation distance from the sample to the detector. Note that this equation is valid in the x-ray regime due to the small diffraction and refraction angles involved with such light.

This linear relationship between the local displacement $\mathbf{v}$ of the modulation features by comparison to a reference position on one side and the phase gradient induced by the sample on the other provides a simple and common way of accessing beam phase information. This core principle is, for instance, applied within the grating interferometer \cite{weitkamp2005} and coded aperture \cite{olivo2007} approaches. Here, the use of near-field speckle as a wavefront modulator is particularly beneficial in the hard x-ray regime because such a pattern is shown to be invariable upon propagation over distances that can range from millimeters to meters \cite{cerbino2008}. Furthermore, this kind of speckle is easily accessible even with beams presenting low transverse coherence properties.

The distinctive aspects of each speckle scheme lie in the approach employed for the recovery of $\mathbf{v}(x,y)$. Figure \ref{fig:methods} illustrates the data that are collected and used within each method.

Within XST (Fig. \ref{fig:methods}.(a)) \cite{berujon2012prl,morgan2012}, the scattering object is fixed and only two images are necessary: one with and one without the sample present in the beam path. The processing consists of taking a small subset of pixels containing some speckle pattern and tracking its shift from the sample image position across the reference speckle image using a cross-correlation algorithm. The operation is repeated for all subsets centered on each pixel. Thus, the vector map $\mathbf{v}(x,y)$ is calculated with subpixel accuracy and the two orthogonal differential wave-front gradients are recovered. Generally, the size of the tracked subset can range from $5\times5$ pixels to more than $21\times21$ pixels. Here the resolution is limited by the grain size of the scatterer, as the superimposition of the speckle grains onto sample features of equivalent size renders the tracking of the modified speckle pattern unstable. Despite the existence of algorithms that take into consideration the effect of subset distortion to enhance the tracking robustness \cite{berujon2015josr}, the method gets, locally, less robust and/or accurate in the presence of a strong phase shift, e.g. at the sample boundaries.

With XSS \cite{berujon2012pra}, whose principle is shown in Fig. \ref{fig:methods}.(b), the scattering object is mounted on a highly accurate bidirectional translation stage in order to raster scan the scatterer transversally to the x-ray propagation direction. Two sets of images are then collected by performing the same scan, whose step is on the order of 100 nm to a few micrometers, with the sample present in the beam and without it (reference). For each pixel, as marked by an eye symbol in Fig. \ref{fig:methods}.(b), two data arrays are collected and the analysis is done pixel by pixel. A cross-correlation-based treatment computes the shift $\mathbf{v}(x,y)$ between the recorded patterns with a substep accuracy.  Finally, the differential phase gradient is retrieved after linear normalization by a geometrical factor \cite{berujon2015}. This scheme can offer a very high spatial resolution when combined with a strongly magnifying setup. Moreover, since the definition of the wave-front gradient is here proportional to the step size employed, it is possible to reach single nanoradian accuracy.  On the other hand, such performance is accessible only at the cost of many sample exposures due to the necessity of performing 2D mesh scans, or at least, two wide-range 1D scans \cite{berujon2012pra}).

XSVT is the most recently introduced speckle approach \cite{berujon2015}. It requires the mounting of the scattering object on a moving stage capable of high repeatability with respect to the effective pixel size of the detector upon geometrical demagnification brought by the setup. However, unlike XSS, no accurate calibration on the motor step size is required. Figure \ref{fig:methods}.(c) shows the stacking up of images when the sample is present and when it is removed from the beam. The stage repeatability is needed for the scatterer to be in exactly the same position when recording images with and without a sample. Under this condition, the data-acquisition procedure consists of recording each $\eta^{th}$ pair of sample and reference images of the two stacks with the scattering object located at every given position of the scatterer. It is also necessary that a fixed pair of images possesses locally a speckle pattern that is uncorrelated from those of the other images. Data analysis starts by building two speckle vectors for each pixel using the two image stacks. Then, each vector from the sample stack is tracked across those of the reference stack. The location of the peak of maximum correlation between the sample’s tracked vectors and the reference ones provides a mean of recovering the local displacement vectors $\mathbf{v}(x,y)$ and hence the differential phase gradient induced by the sample. This protocol means that a resolution close to that of the detector can be reached and that the sensitivity scales with the detector pixel size and the propagation distance.

\begin{table}
\begin{tabular}{|l||c|c|c|}
  \hline
  ~ & XST& XSS& XSVT\\
  \hline\hline
  Specific setup & None & Accurate& Repeatable\\
  requirement & ~& motor &motor\\
 \hline
  Sample exposure & 1 & $\mathcal{O}(\eta^2)$ & $\mathcal{O}(\eta)$\\
  \hline
  Sensitivity & $\sim ~ 0.1~\mu$rad & $\sim ~  0.001~\mu$rad & $\sim ~ 0.01~\mu$rad\\
  \hline
  Spatial  & $\sim ~ 10~\mu m$ & $\sim ~ 0.1~\mu m$& $\sim ~ 1~\mu m$ \\
  resolution &(~grain size) &(~pixel size) & ($\geq$pixel size)\\
  \hline
\end{tabular}
\caption{Comparative table of speckle based techniques.}
\label{tab:comp}
\end{table}

Table \ref{tab:comp} presents a brief survey of the attributes of each of the three techniques. Note that the figures provided are discussed in more detail elsewhere \cite{berujon2015josr,berujon2013apl,berujon2012prl,berujon2015,berujon2012pra,wang2015a,berujon2013phd}. We recognize that, in all three methods, the dark-field signal can be obtained by considering the change in speckle visibility between the references and sample images. However, although this capability is demonstrated with XST \cite{zanette2014}, only the XSS and XSVT approaches presently prove to be quantitative and accurate enough for 3D tomographic reconstruction. The moderate number of images per projection necessary for XSVT to access the 2D differential phase gradient in addition to the dark-field signal constitutes the main asset of the technique. The data redundancy in the XSVT technique permits the sensitivity to be improved down to 10 nrad. From these considerations and from Table \ref{tab:comp} we are able to conclude that the XSVT method represents the best compromise on the number of necessary images per projection with regard to the sensitivity and spatial resolution. Therefore, although tomography is demonstrated with XST \cite{wang2015,zanette2015}, XSVT represents our approach of choice to achieve high-resolution phase-contrast tomography at a cost-effective number of sample exposures. Nevertheless, one can foresee that XSS will be able, in the future, to perform 3D tomographic imaging of samples presenting very small electronic variations when the dose is not an issue.

\section{Experimental setup}

\begin{figure}[htb!]
\includegraphics[width=\columnwidth]{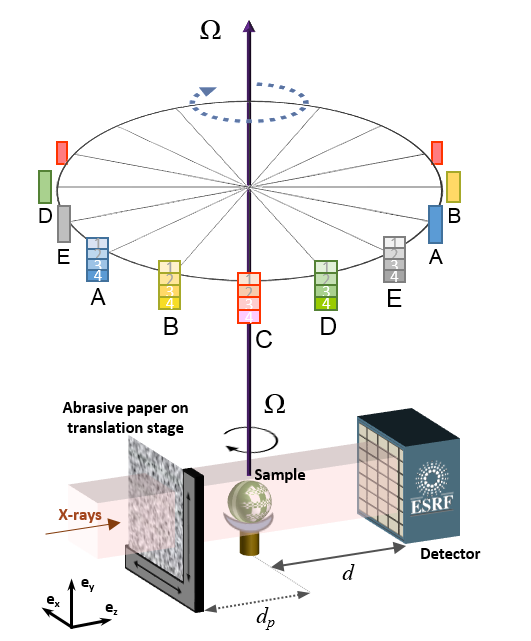}
\caption{Bottom: setup of the experimental arrangement. Top: acquisition scheme used for the data collection. For each angular projection, a set of four images images (either A, B, C, D or E) are collected with the membrane located at distinctive positions. Each of these sets of 4 different speckle illuminations are reused every five projection angles, meaning that in total 20 independent speckle illumination are used.\label{fig:setup}}
\end{figure}

A sketch of the experimental setup is shown in Fig.\ref{fig:setup}. The experiment is conducted at the ESRF beamline BM05 \cite{ziegler2004}, where a precision tomography station is permanently installed. At BM05, the photons are produced by a 0.85 T bending magnet on the circulating 6.02 GeV electrons of the storage ring. The photon energy is narrowed down to a band of $\Delta E /E \sim 10^{-4}$ centered around 17 keV by means of a double-crystal Si(111) monochromator. The sample is placed at 55 m from the source and the detector, a CCD-based FReLoN (Fast Read-out Low Noise) camera coupled to a scintillator, $d=1$ m further downstream from the sample. The effective pixel size of the imaging system is $h=5.8$ $\mu$m. A piece of abrasive paper with a grit designation of P800 is mounted on a linear translation stage allowing displacements in the beam transverse directions and located $d_p=$ 400 mm upstream of the sample. At a distance of $d= 1$ m from the sample, denoting by$<~>$ and $\sigma$, respectively, the mean value and the standard deviation of the vector scalar components, the visibility of the speckle presenting a nearly Gaussian distribution is of $(I_{max}-I_{min})/2<I> = 0.4$, or $\sigma(I)/<I> = 0.13$. The recorded images have a size of 2048 pixels horizontally by 1850 pixels vertically.

The selected samples presented here are juniper berries and red currant berries. The juniper sample is interesting for the presence of features at very different spatial frequencies and for the significant amount of x-ray scattering generated by structures at the submicron scale. By its very nature the juniper berry is a difficult sample to image for many techniques. In comparison the red currant presents fewer parts that generate strong x-ray scattering but has large homogeneous features that can sometimes become problematic as they render less-accurate quantitative results with propagation-based methods.

\section{XSVT-based tomography with an interlaced scheme\label{sec:m1}}

The method of structured illumination employed in this section has been quickly presented in Sec. \ref{sec:schemes} and is detailed for projection imaging Ref.~\cite{berujon2015}. Here in addition, an interlaced scheme is used to optimize the number of images collected for each projection. The proposed interlaced XSVT method allows us to greatly reduce the overall number of sample exposures whilst conserving the accuracy of the phase and dark-field signals by factoring in the images recorded at consecutive projection angles.

The data-acquisition scheme is sketched at the top of Fig.~\ref{fig:setup}: a total of 20 distinctive transverse positions of the membrane with respect to the beam direction are defined and split in five sets $p=[A,B,C,D,E]$. For each of the $N = 1800$ angular projections over a 180-degree tomography scan, four images are recorded with the membrane located at the positions $k_p=1,2,3,4$ defined for each angular set. For each projection angle $\Omega_n$ with $n \in |[1,N]|$, these positions are defined as the modulo 5 of the projection number thus generating a cyclic repetition of the set $p$ employed.

Experimentally, the full data set results from the collection of 20 different scans of 360 projections, the starting scan angle being shifted by 0.1 degree every four scans. Then, for each projection $n$, the interlaced scheme consists of using the image intensity $i(\Omega_n,x,y,k_p)$ recorded at the projection $\{n-2,n-1,n,n+1,n+2\}$ as a multivariate random variable to build speckle vectors $\mathbf{s_v}(\Omega_{n},x,y)$ of size 20:

\begin{equation}
\begin{aligned}
\mathbf{s_v}(\Omega_{n},x,y) =  \bigg( & \Big\{i(\Omega_{n-2},x,y,k_p)\Big\}_{k_p} ,...,\\
 & \Big\{i(\Omega_{n},x,y,k_p)\Big\}_{k_p},...,\\
 & \Big\{i(\Omega_{n+2},x,y,k_p)\Big\}_{k_p} \bigg)
\end{aligned}
\end{equation}
with the sets denoted $\{~\}$ built with $k_p=1,..,4$.

\begin{figure*}
\centering
\includegraphics[width=0.85\textwidth]{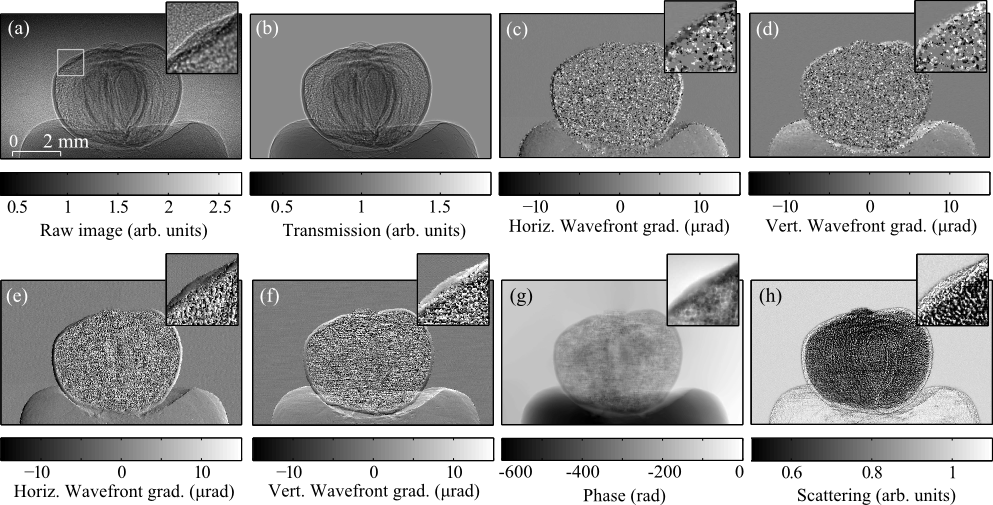}
\caption{Images of the juniper-berry sample. (a) A raw image of the juniper-berry sample at the first angular projection with an inset showing the speckle modulation of the wave-front. (b) The corresponding calculated absorption image. (c) Horizontal and (d) vertical wave-front gradients calculated with the XST technique \cite{berujon2012prl}. (e) Horizontal and (f) vertical wave-front gradients calculated with the interlaced XSVT method. (g) Phase image from the 2D integration of (e-f). (h) Dark-field image obtained with the XSVT interlaced scheme. Insets of the images are defined by the zone in (a) marked out with a white square.
\label{fig:images}}
\end{figure*}

Such an interlaced scheme ensures the avoidance of any redundancy in the speckle vectors that would otherwise decrease the amount of independent statistics they contain and then reduce the accuracy of the vector correlation calculation \cite{berujon2015}. Another two sets of images are collected below the 0 degree and beyond the 180 degree positions in order to complete the data sets of the first, second, penultimate, and last projections.

\begin{figure}
\includegraphics[width=0.85\columnwidth]{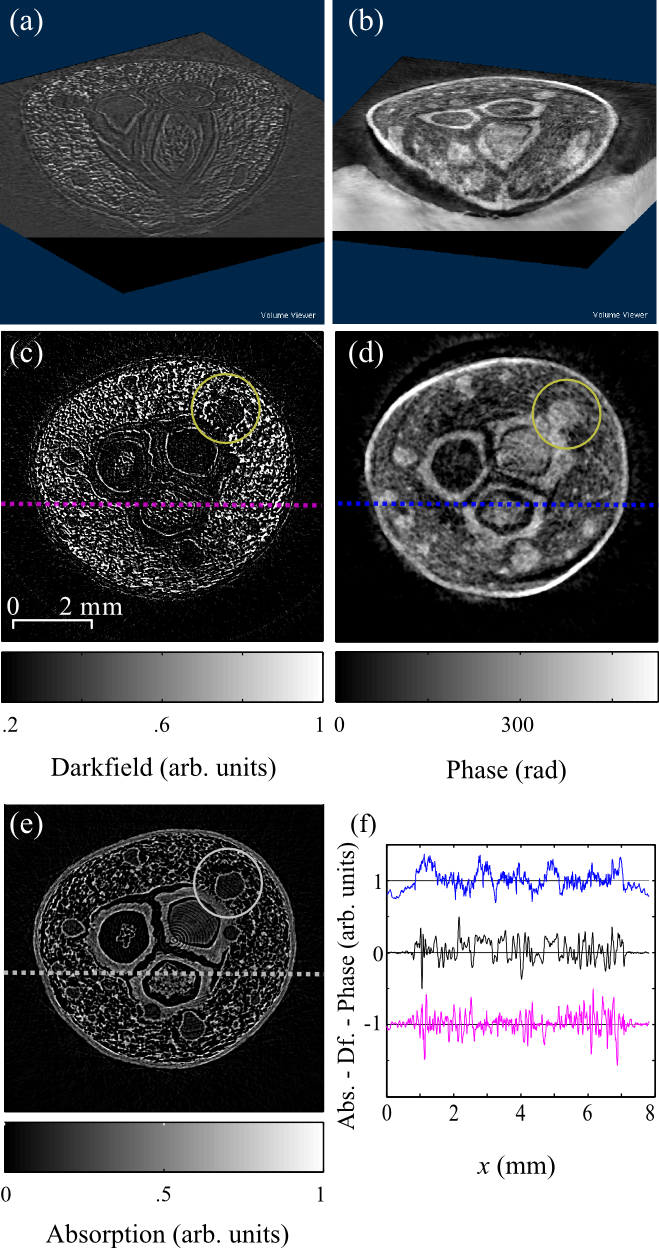}
\caption{The cut in the volume reconstructions of the juniper-berry (a) from the phase signal and (b) from the dark-field signal. (c) An example horizontal transverse slice reconstructed from the dark-field signal and the same slice calculated with (d) the absolute phase shift signal and (e) the absorption signal. (f) Cuts along the intensity lines shown in (c,d,e) after shift and scaling for comparison purposes. \label{fig:slices}}
\end{figure}

A reference image $g(p,x,y,k_p)$, i.e., without the sample inserted into the beam, is taken before each 20 of the scans. Hence, a total of five reference matrices of size $2048\times1850\times20$ could be built by a circular permutation of the speckle-vector elements and used as reference vectors $\mathbf{r_v}(\Omega_n,x,y)$:

\begin{equation}
\begin{aligned}
\mathbf{r_v}(\Omega_n,x,y) =  \bigg( & \Big\{g(\zeta^q(A),x,y,k_p)\Big\}_{k_p} ,...,\\
 & \Big\{g(\zeta^q(C),x,y,k_p)\Big\}_{k_p},...,\\
 & \Big\{g(\zeta^q(E),x,y,k_p)\Big\}_{k_p} \bigg)
\end{aligned}
\end{equation}

where $\zeta$ is the elementary cyclic permutation over the elements of $p$ with order $q = \mod(n-1,5)$.
From these speckle vectors of length 20, the absorption and dark-field signals in addition to the two differential phase gradients could be calculated for each $\Omega_n$ using the XSVT method sketched in Fig. \ref{fig:methods}.(c) and detailed in Ref. \cite{berujon2015}.

In concrete terms, the displacement of the speckle vectors $\mathbf{v}(\Omega_n,x,y)$ induced by the sample refraction is calculated using the Pearson correlation coefficient $\rho$:

\begin{equation}
(v_x,v_y) = - \argmax_{(v_X,v_Y)}\rho(\mathbf{s_v}(x,y),\mathbf{r_v}(x+v_X,y+v_Y))\label{eq:disp}
\end{equation}

with $\mathbf{v} (x,y)=v_x\mathbf{e_{x}}+v_y\mathbf{e_{y}}$ and where the variable $\Omega_n$ is dropped for clarity. Note that the vectors $\mathbf{v}$ are calculated with subpixel accuracy \cite{pan2009}.

Next, the differential phase $\nabla \phi$ is obtained using the formula already mentioned above and valid for the small angles:

\begin{equation}
\nabla \phi(x,y) = \frac{2\pi}{\lambda}\frac{\mathbf{v}(x,y)}{d}
\label{eq:phase}
\end{equation}

where $\lambda$ is the photon wavelength. We call the wave-front gradient $\nabla W = \mathbf{v}(x,y)/d$, which corresponds to the angular deflection of the rays caused by refraction. The transmission signal $T$ for each projection is calculated by taking the ratio of the average speckle- and reference-vector intensity elements:

\begin{equation}
T(x,y) = \frac{<\mathbf{s_v}(x+v_x,y+v_y)>}{<\mathbf{r_v}(x,y)>}
\label{eq:trans}
\end{equation}

Besides, the dark-field signal $D_f$ is calculated by using:
\begin{equation}
D_f(x,y) = \frac{\sigma(\mathbf{s_v}(x+v_x,y+v_y))}{ <\mathbf{s_v}(x+v_x,y+v_y)>}\frac{<\mathbf{r_v}(x,y)>}{\sigma(\mathbf{r_v}(x,y))}
\label{eq:dark}
\end{equation}

where $\sigma$ denotes the standard deviation of the vector elements.

Note that this last formula is based only on statistics. Hence, in the case of large vectors ($\mathbf{r_v}$,$\mathbf{s_v}$), the approach becomes also valid in the case where $\mathbf{s_v}$ and $\mathbf{r_v}$ are built from images with different illuminations, i.e. with the membrane located at uncorrelated transverse positions between the reference and sample image stacks \cite{boas2010}.

The phase-imaging method employed here is equivalent to x-ray grating interferometry or analyzer-based imaging as the measured signal is the differential phase signal $\nabla \phi$ \cite{momose2006,pfeiffer2007,gasilov2014}. In these methods, a specific filter is often used in the inverse Radon transform to account for the specificity of the signal and perform a one-dimensional integration in the Fourier space \cite{nesterets2006,pellica2015}. Here, as the XSVT technique provides the two transverse differential phase maps, in contrast to the previous techniques, the phase images can be recovered by 2D integration by matrix inversion via, for instance, the Cholesky decomposition.

A set of raw, directional phase gradient, phase and dark-field images is shown in Fig. \ref{fig:images}.(a-h) for the first angular projection of the juniper-berry sample. Figure \ref{fig:images}.(c-f) makes us realize that whilst the traditional XST fails largely to track the speckle subsets due the presence in the sample of high spatial frequency features, the XSVT method succeeds perfectly to sense the very turbid wave-front. With the proposed interlaced XSVT technique, both the vertical and horizontal phase gradients are correctly recovered for each pixel, which means that upon 2D integration, the correct phase images are conserved despite the presence of fast-varying structures as seen in the inset of Fig. \ref{fig:images}.(g). Similarly, the statistics offered by the speckle vectors permits a clear sensing of the dark-field signal as seen in Fig. \ref{fig:images}.(h).

From the three imaging modalities, 3D reconstructions are operated using the ESRF PyHST (Python High Speed Tomography) \cite{pyhst} implementation with the classic filtered back-projection algorithm. Volume reconstructions from the phase and dark-field signals are shown in Fig. \ref{fig:slices}.(a-b). Transverse reconstruction slices are also displayed for comparison in Fig. \ref{fig:slices}.(c-e). These three signals do not render the same contrast for numerous sample features. For instance, the scattering features in the outer part of the juniper-berry (noted 1) render a strong contrast in the dark-field signal (slice in Fig. \ref{fig:slices}.(c)) while a low contrast is observed in the phase slice (Fig. \ref{fig:slices}.(d)). Similarly, within the slice of the phase signal, homogeneous features observed within the yellow circle exhibit a large phase shift while presenting very low absorption. The comparison cuts in Fig. \ref{fig:slices}.(f) provide a different view on these results.

While the total number of images collected (7200) may seem large, it could be readily diminished by reducing the number of projections in the angular tomography scan. This figure could also be compared with holotomography for instance, which also requires a minimum of four images per projection or with grating interferometry, which demands phase-stepping scans of ~four steps per projection. For this latter technique, although a special scheme could slightly reduce the number of images necessary for reconstruction \cite{zanette2011}, the presence of an absorption grating is unfavorable when the irradiation dose received by the sample is an issue. The presented scheme can also benefit from a smaller ensemble of images per projection by using images of further projections in order to maintain the length and then the amount of statistics contained in the speckle vectors.

\section{A mixed XST-XSVT approach\label{sec:m2}}

\begin{figure}
\includegraphics{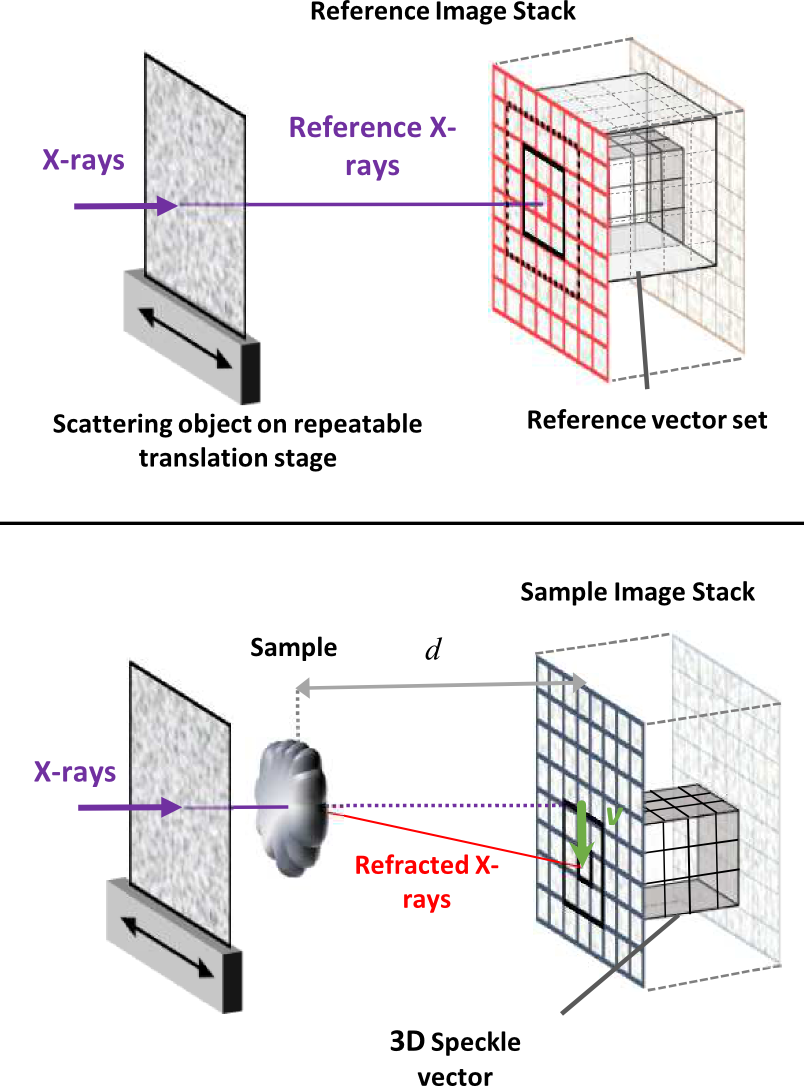}
\caption{Mixed XST-XSVT data-processing scheme. Therein, the pixel values considered for the construction of the speckle vectors are marked out with black and correspond to the location of the central pixel. For the construction of the vectors, values of the closest surrounding pixel are used in addition to the central pixel value of the five images. \label{fig:vectmeth2}}
\end{figure}

We present in this section a second scheme for phase sensing and tomography reconstruction with the aim of reducing the number of acquisitions per projection even further. The principle of this method consists of a mixed scheme between the traditional XST technique \cite{berujon2012prl,wang2015} and the optimized XSVT scheme presented in the previous section. The experimental protocol consists of operating the data acquisition according to the procedure sketched in Fig. \ref{fig:setup} (top), while recording only one image per projection angle, i.e. $k_p=1$. This means that during the scan the light is now modulated with five different speckle patterns, with a cyclical repetition of every five angular projections.

As done earlier, the five interlaced images centered on each considered projection are used to build a vector with sufficient statistics to achieve an accurate tracking. Furthermore, we enlarge the vectors by using not only the values of the centered pixels of each image but also the ones of the eight surrounding pixel values as displayed in Fig. \ref{fig:vectmeth2}. This approach permits to generate, in our case, vectors containing 35 components:

\begin{equation}
\begin{aligned}
\mathbf{s_v}(x,y) =  \bigg( &  \Big\{i(\Omega_{n-2},x+\varepsilon_x,y+\varepsilon_y)\Big\}_{(\varepsilon_x,\varepsilon_y)} ,..., \\
 & \Big\{i(\Omega_{n},x+\varepsilon_x,y+\varepsilon_y)\Big\}_{(\varepsilon_x,\varepsilon_y)} ,...,\\
 & \Big\{i(\Omega_{n+2},x+\varepsilon_x,y+\varepsilon_y)\Big\}_{(\varepsilon_x,\varepsilon_y)} \bigg)
\end{aligned}
\end{equation}
with here $(\varepsilon_x,\varepsilon_y) \in \{-h,0,h\}^2$, $h$ being the pixel size. Reference stacks are built in the same manner from the reference images:
\begin{equation}
\begin{aligned}
\mathbf{r_v}(\Omega_n,x,y) =  \bigg( & \Big\{g(\zeta^q(A),x + \varepsilon_x,y+\varepsilon_y)\Big\}_{(\varepsilon_x,\varepsilon_y)} ,...,\\
 & \Big\{g(\zeta^q(C),x+\varepsilon_x,y+\varepsilon_y)\Big\}_{(\varepsilon_x,\varepsilon_y)},...,\\
 & \Big\{g(\zeta^q(E),x+\varepsilon_x,y+\varepsilon_y)\Big\}_{(\varepsilon_x,\varepsilon_y)} \bigg)
\end{aligned}
\end{equation}

The following step of the processing method is the one of XSVT described by Eq.~\ref{eq:disp}-\ref{eq:dark} applied to these new vectors.

\begin{figure}
\includegraphics{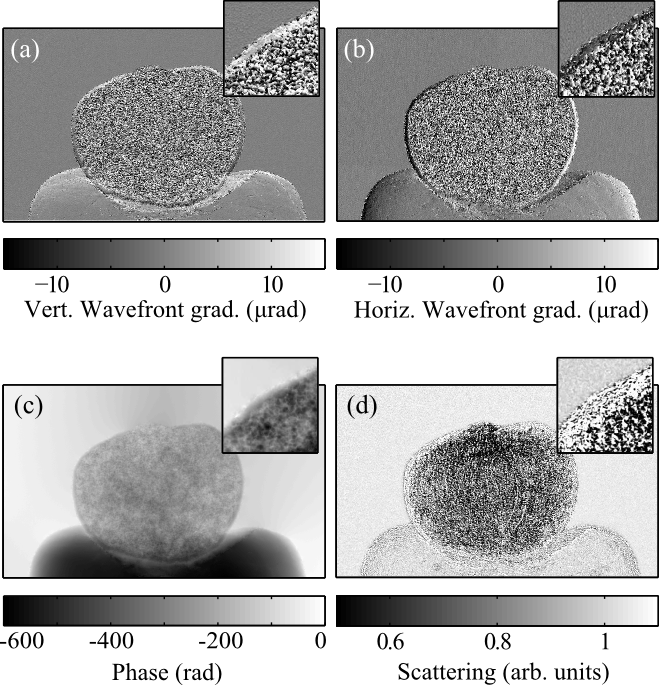}
\caption{Images of the juniper-berry sample obtained with the mixed XST-XSVT approach. (a) Vertical and (b) horizontal wave-front gradients. (c) Phase and (d) dark-field images. \label{fig:imGen2}}
\end{figure}
\begin{figure*}
\includegraphics[width=0.85\textwidth]{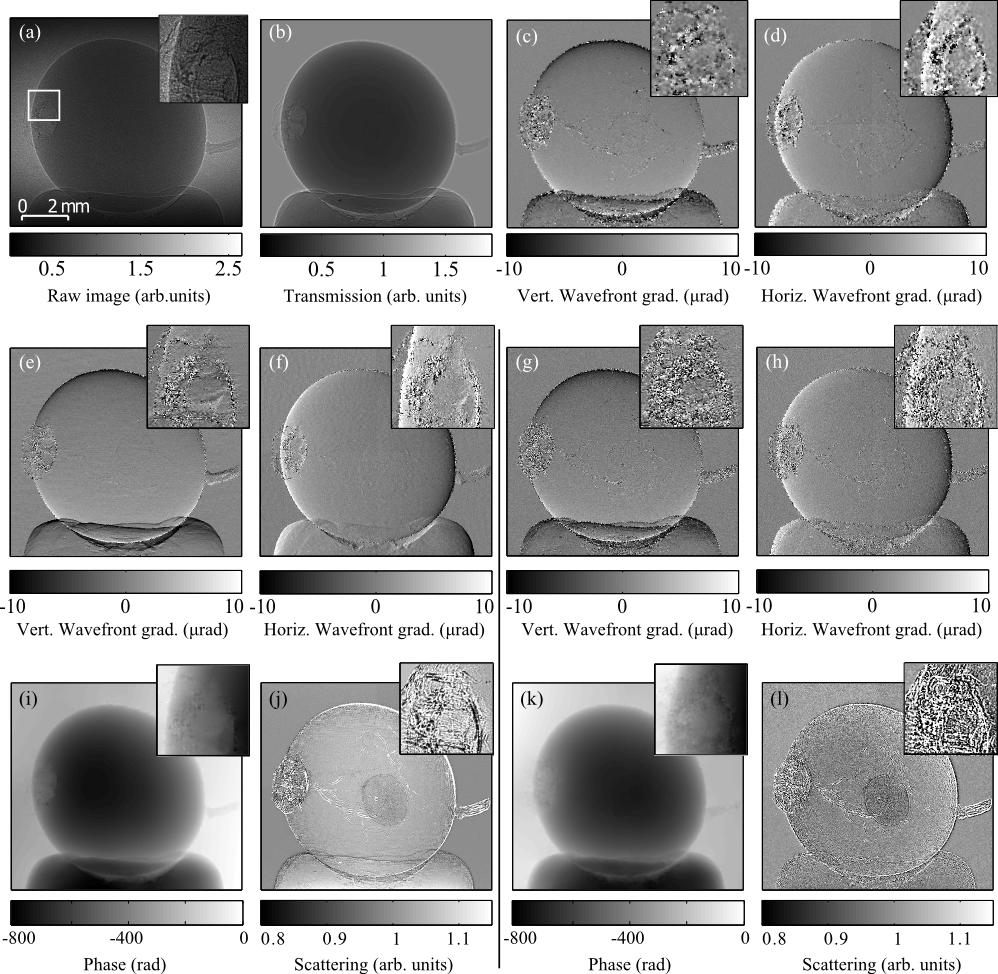}
\caption{Images of the red-currant sample. (a) Raw image of the sample at the first angular projection with an inset showing the area marked out with a white square. (b) The corresponding calculated absorption image. Vertical and horizontal wave-front gradients calculated with: (c-d) the XST technique, (e-f) the interlaced XSVT method, (g-h) the mixed XST-XSVT approach presented in this section. (i) Phase and (j) dark-field images extracted with the interlaced XSVT method. (k) Phase and (l) dark-field images extracted with the mixed XST-XSVT approach.\label{fig:gros2}}
\end{figure*}

The calculated differential phase and phase images obtained with the mixed XST-XSVT technique are shown in Fig. \ref{fig:imGen2}.(a-d) for the juniper sample. Figure \ref{fig:gros2} shows the equivalent calculated maps for the red-currant sample. Images (e-f, i-j) are extracted with the interlaced XSVT technique previously presented while images on the right side, labeled (g-h, k-l), are the maps calculated with the present mixed XST-XSVT technique. From these two sample projection images, although the two methods apparently render similar results, a meticulous observation shows that the images obtained with the mixed XST-XSVT method reveal less detail. This difference is better emphasized when comparing the insets of the wave-front gradient maps. In the juniper-berry sample, some features are smeared out with the mixed technique, thus preventing a correct three dimensional reconstruction. Likewise, the scattering image obtained with the mixed method offers less signal-to-noise ratio because the statistics required for its calculation have less independent data.

\begin{figure*}
\includegraphics[width=0.85\textwidth]{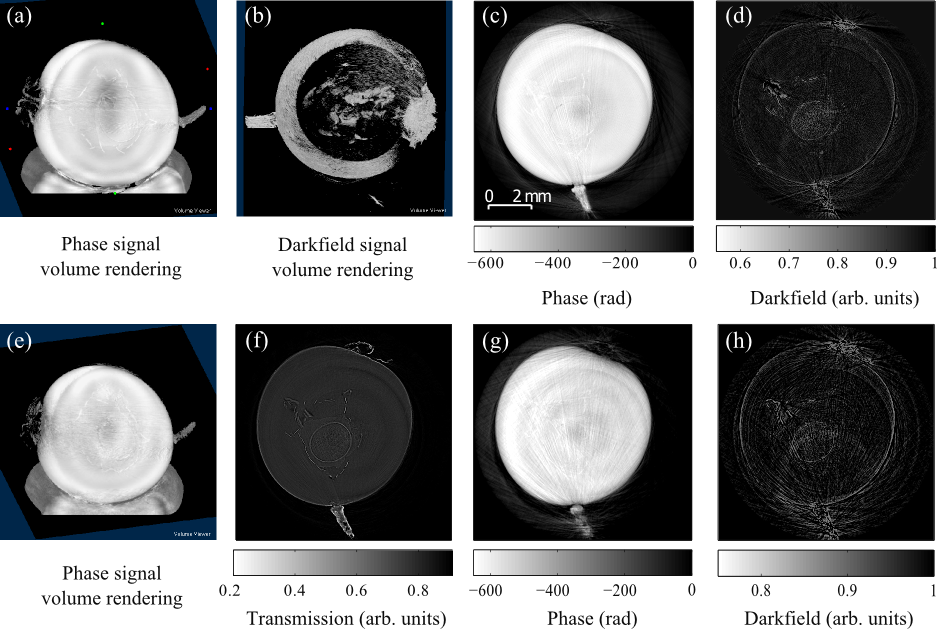}
\caption{Volume rendering of the tomographic reconstructions of the red-currant sample (a) from the phase signal and (b) from the dark-field signal extracted with the XSVT method of Sec. \ref{sec:m1}. (c) Slice reconstruction from the phase signal and (d) from the dark-field signal using the same method. (e) Volume rendering of the reconstruction from the phase signal extracted with the mixed XST-XSVT method. Slice reconstructions obtained with this same method: (f) absorption signal, (g) phase signal and (h) dark-field signal. \label{fig:slicem2}}
\end{figure*}

Conversely, the red-currant sample, which generates much less scattering than the juniper-berry sample, could be easily reconstructed by using the classic filtered-back-projection algorithm from the different signals extracted with this mixed approach. Reconstruction samples are shown in Fig. \ref{fig:slicem2}. This example demonstrates the correctness and greater appropriateness of this approach over the traditional XST technique which, in this particular example, failed to accurately 3D reconstruct the smaller features.

More generally, the results presented in Figs. \ref{fig:images}, \ref{fig:imGen2}, \ref{fig:gros2} and \ref{fig:slicem2} demonstrate a better fidelity of the signal recovery with the interlaced XSVT scheme than for the other speckle approaches. In comparison to the traditional XST approach \cite{wang2015}, the mixed scheme processing is highly beneficial to the robustness of the phase sensing. Where fast and sharp variations of the sample electronic density generate wave-front features with sizes comparable to the speckle, the direct XST of speckle subsets becomes quickly limited and/or the spatial reduction drastically reduced. In this regard, the interlaced mixed method offers much higher efficiency and reliability albeit at the cost of a slight degradation in spatial resolution with respect to the XSVT method.

This second scheme, integrating both multi-image and multipixel values together in a single vector, allows the statistics necessary for accurate correlation and tracking of the speckle displacements to be improved. Nevertheless, such artificially enlarged vectors do not improve the statistics as much as one would expect by a straightforward calculation based on the ratio of integrated pixels (here nine). The fact that the values of the neighboring pixels are partially correlated results in a statistics gain lower than the number of pixels considered. This can be illustrated by comparing the sensitivity of the two techniques when measuring the standard deviation of the wave-front gradient maps in areas of $100\times100$ pixels in the absence of a sample in the x-ray beam. For the XSVT method of Sec. \ref{sec:m1}, the measured standard deviation is of $\sim$0.35 $\mu$rad, while it is of $\sim$0.75 $\mu$rad for the mixed XSVT-XST method.

As no additional images are required within this processing scheme compared to the XST technique or propagation based methods, this approach presents a high potential for the imaging of samples for which the dose is a sensitive issue.

\section{Conclusion}

In conclusion, we demonstrate that the XSVT technique is suitable for high-resolution absorption, phase and dark-field contrast tomography. The several images collected for each projection permit us to overcome the limited spatial resolution of the XST technique, whilst the interlaced scheme and the use of a nonabsorbing scattering membrane allows one to keep the dose at a moderate level. Furthermore, the alternative processing scheme presented in the second part is a very promising way of improving the performance of the XST technique in terms of robustness and spatial resolution. Overall, the XSVT method is expected to find many applications at laboratory sources, synchrotron, and also Compton-accelerator-based x-ray sources, for samples where the dose and the image resolution are a problematic combination.

\begin{acknowledgments}
The authors thank V. Fernandez for help in the handling of the BM05 beamline tomography stage, L. Graham for the review of the manuscript and the ESRF for the financial support.
\end{acknowledgments}

\bibliography{BibIntlTomo}

\end{document}